\documentclass[english,a4paper,12pt]{article}
\usepackage{babel,amsmath,amsthm,graphics,graphicx,eucal,amscd,amsfonts,amssymb}
\usepackage[small]{caption}
\newtheorem{theorem}{Theorem}[section]
\newtheorem{lemma}{Lemma}[section]

\newcommand{\Zset}{\mathbb{Z}}
\def \dis{\displaystyle}

\newcommand{\q}{\tilde{q}}
\def \n{\mspace{-15.0mu}}

\begin{document}

\begin{center}
{\Large \textbf{On localized solutions of discrete nonlinear Schr\"{o}dinger equation. An exact result.}}\\
\vspace{1cm}
{\large P. Pacciani $^1$, V. V. Konotop $^2$, G. Perla Menzala $^3$}\\
\vspace{.5cm}
{
\small \em
$^1$ Centro de F\'{\i}sica
Te\'orica e Computacional, Universidade de Lisboa, Complexo Interdisciplinar, Av. Prof. Gama Pinto, 2, Lisboa 1649-003, Portugal\\
$^2$ Centro de F\'{\i}sica
Te\'orica e Computacional, Universidade de Lisboa, Complexo Interdisciplinar, Av. Prof. Gama Pinto, 2, Lisboa 1649-003, Portugal and Departamento de F\'{\i}sica, Faculdade de Ciências,Universidade de Lisboa, Campo Grande, Ed. C8, Piso 6, Lisboa 1749-016, Portugal\\
$^3$ National Laboratory of Scientific Computation,
Av. Getulio Vargas 333, 25651-070 Petropolis, C.P. 95113, RJ Brazil
}
\end{center}

\vspace{1cm}

\noindent\rule{13.5cm}{0.01cm}

\begin{abstract}
Local and global existence of localized solutions of a discrete nonlinear Schr\"{o}\-ding\-er (DNLS) equation, with arbitrary on-site nonlinearity, is proved. In particular, it is shown that an initially localized excitation  persists localized during infinite time. Moreover, if initial localization is stronger than $|n|^{-d}$  with any power $d$, it maintains itself as such during infinite time. The results are generalized to various types of inter-side and saturable nonlinearities, to lattices with long range interactions, as well as DNLS with dissipation.
\end{abstract}


\vspace{.5cm}
\noindent\rule{13.5cm}{0.01cm}

\section{Introduction}

Localization and transfer of energy in nonlinear lattices are the key problems of the physics of discrete systems, which have attracted a great deal of attention during the last years (see e.g. \cite{Vaz}). In this context, the discrete nonlinear Schr\"{o}dinger (DNLS) equation
\begin{eqnarray}
\label{Seq1}
i\dot{q}_{n}(t)+\Delta q_n(t)-\chi |q_n(t)|^{2(p-1)}q_n(t)=0\ ,
\end{eqnarray}
where $p$ is an integer which throughout this paper will be considered bigger than or equal to 2, $p\geq 2$, $\chi=\pm 1$, and $\Delta q_{n}=q_{n+1}+q_{n-1}-2q_{n}$, is an object of special interest~\cite{review} due to its numerous physical applications, ranging from dynamics of electrons in one-dimensional crystals in the tight-binding approximation, to arrays of optical waveguides, with Kerr ($p=2$) and non-Kerr ($p\geq 3$) nonlinearities, and arrays of Bose-Einstein condensates.

A possibility of energy localization in the form of so-called intrinsic localized modes, i.e. periodic in time and  strongly localized in space solutions  of Eq. (\ref{Seq1}), was predicted in Ref.~\cite{ILM}. Later on, in Ref.~\cite{McKay1994}, this fact received its mathematical proof, and recently a classification of diversity of intrinsic localized modes has been provided~\cite{ABK}.
By its definition, an intrinsic localized mode can be viewed as a particular solution of the Cauchy problem for the DNLS equation, which has sufficiently large amplitude and strong spatial localization, thus belonging to a subclass of all possible solutions.  Meanwhile, the problem of localization properties of an arbitrary initially localized pulse, remains open. This justifies one of the main objectives of the present paper: to provide a rigorous proof of persistence in time of localization of solutions of the Cauchy problem for the DNLS equation (\ref{Seq1}) subject to an initial condition
\begin{equation}
\label{Seq1_1}
q_n(0)=a_n\ ,
\end{equation}
where $a_n$, with $n\in\mathbb{Z}$, are given complex constants. Initial localization of the pulse will be characterized by the constraint
\begin{equation}
\label{hp}
A_d=\sum_{m=-\infty}^{+\infty}(1+m^2)^d|a_m|^2<+\infty\ ,
\end{equation}
where $d$ is a positive integer, $d\geq 1$, which characterizes the decay of $|a_m|$ when $|m|$ tends to infinity. If (\ref{hp}) is verified for all $d\leq d_0$ and fails for $d\geq d_0$ the localization is referred to as power. As it is clear, for modes localized {\em exponentially} (which is also the case of intrinsic localized modes mentioned above), (\ref{hp}) holds for all positive $d$.

Another important issue related to the DNLS model (\ref{Seq1}) is that it can be viewed as a discretization of the nonlinear Schr\"{o}dinger equation with high nonlinearity. As it is known~\cite{Sulem}, such equation at a given sign of $\sigma$ possesses blowing up solutions. Collapse occurs due to small wave-lengths (large frequencies). Thus it is natural to expect that the discretization, which introduces the linear spectrum frequency cut-off, prevents collapse in (\ref{Seq1}) with any integer $p$ and both $\chi=\pm 1$, which indeed was observed in \cite{Bang}. In the present paper consideration of the localization properties of solutions of the DNLS equation will be reduced to the existence theorems and thus arresting collapse will be automatically guaranteed (we notice that this approach is analogous to one used in \cite{MK} for nonlinear Klein-Gordon and sine-Gordon lattices).

The organization of the paper is as follows. In Sec.~\ref{sec:descrip}, we describe the statement of the problem, define an appropriate Banach space,  and formulate the main result. Sec.~\ref{sec:local} is devoted to the proof of the local existence of localized solutions. Global existence is proved in Sec.~\ref{sec:global}. In Sec.~\ref{dissip} the results are generalized to the DNLS equation with dissipation, as well as models with long-range interactions and other types of nonlinearity. The results are summarized in Conclusion.

\section{Statement of the problem and main results}
\label{sec:descrip}

By analogy with~\cite{MK} our proof will be based on the {\em Banach fixed point theorem}~\cite{kant} (see also below). For its implementation we rewrite the nonlinear model (\ref{Seq1}) in an integral form. To this end we first consider the linear problem associated with (\ref{Seq1}):
\begin{equation}
\label{Seq-lin}
i\dot{p}_{n}(t)+\Delta p_{n}(t)=0 \ ,\qquad
               p_n(0)=a_n \ ,
\end{equation}
which solution is trivially obtained  in the form
\begin{equation}
\label{Seq-p}
p_n(t)=\sum_{m=-\infty}^{+\infty}a_m G(n-m,t)\ ,
\end{equation}
with $G(n,t)$ being the Green function
\begin{equation}
\label{Green}
G(n,t)=\frac{1}{\pi}\int_{-\frac{\pi}{2}}^{\frac{\pi}{2}} f(\sigma,t) d\sigma,\qquad f(\sigma,t)\equiv e^{2i\sigma n-4it\sin^2\sigma}\ .
\end{equation}
Then, using (\ref{Seq-p}) and (\ref{Green}) we rewrite (\ref{Seq1}) as
\begin{equation}
\label{int}
q_n(t)=p_n(t)+\sum_{m=-\infty}^{+\infty}\int_0^tG(n-m,t-s)F\left(q_m(s)\right) d s\ ,
\end{equation}
where  $F\left(q_n(t)\right)$ is given by:
\begin{equation}
\label{F}
F\left(q_n(t)\right)=\chi |q_n(t)|^{2(p-1)}q_n(t)\ .
\end{equation}

One of the main objects of our analysis will be a $2d$-momentum $Q_d(t)$ defined as follows
\begin{equation}
\label{Q}
Q_d(t)=\sum_{n=-\infty}^{+\infty}\left(1+n^2\right)^d\left|q_n(t)\right|^2\ .
\end{equation}
At $d=0$, (\ref{Q}) becomes a trivial integral of (\ref{Seq1}),
\begin{equation}
\label{N0}
Q_0=\sum_{n=-\infty}^{+\infty}\left|q_n(t)\right|^2 ,\qquad \frac{ d Q_0}{ d t}=0 ,
\end{equation}
which often is referred to as a {\em number of particles} $N$ and therefore below we use  notation $Q_0=N$. For an arbitrary $d$ $Q_d(t)$ may grow, but it always remains bounded. Indeed, one can use Cauchy-Schwarz's inequality and compute a derivative of $Q_d(t)$ from (\ref{Seq1}), in order to deduce the estimate as follows:
\begin{eqnarray*}
\frac{ d}{ d t}\,Q_d(t)=2\sum_{n=-\infty}^{+\infty}\left\{\left[1+\left(1+n\right)^2\right]^d-\left(1+n^2\right)^d\right\}\Im(\bar{q}_nq_{n+1})\\
 \leq
C_0\sum_{n=-\infty}^{+\infty}\left(1+\left(1+n\right)^2\right)^{d/2}
\left(1+n^2\right)^{d/2}|q_n||q_{n+1}|
\leq C_0Q_d(t)\ .
\end{eqnarray*}
Here $C_0=[(\sqrt{5}+1)^d-(\sqrt{5}-1)^d]/2^{d-1}$ and it was used
\[
\frac{\sqrt{5}-1}{\sqrt{5}+1}<\frac{1+\left(1+n\right)^2}{1+n^2}<\frac{\sqrt{5}+1}{\sqrt{5}-1}\ .
\]

Then the positive real function $Q_d(t)$ satisfies the hypothesis of Gronwall's lemma and hence
\begin{equation}
\label{ineq-Q}
Q_d(t)\leq Q_d(0)e^{C_0t}\ .
\end{equation}

The next step to be performed, for the sake of the use of the Banach fixed point theorem, is defining the appropriate Banach space. To this end, for some $T>0$, we introduce a function $q(t)$:
\begin{equation}
\label{func_q}
q(t)=\left(\ldots,q_{n-1}(t),q_n(t),q_{n+1}(t),\ldots\right)\ ,
\end{equation}
such that: $q_n(t)$ are continuously differentiable for any $n$, $q_n(0)=a_n$ and 
\[
\dis\sup_{0\leq t<T}\left\{\sum_{n=-\infty}^{+\infty}\left(1+n^2\right)^d\left|q_n(t)\right|^2\right\}<\infty\ ,
\]
and consider the linear space $X^d(T)$, which consists of all such functions $q(t)$. We also define a norm of an element $q(t)\in X^d(T)$ by
\begin{equation}
\label{norm}
\left\|q\right\|_{X^d}^{2}=\dis\sup_{0\leq t<T}\left\{\sum_{n=-\infty}^{+\infty}\left(1+n^2\right)^d\left|q_n(t)\right|^2\right\}\ .
\end{equation}
Then the space $\left(X^d(T),\left\|\cdot\right\|_{X^d}\right)$ becomes a Banach space.

As it is evident, the norm $\|q\|_{X^d}$ can be used as a measure of localization of the solution: if a solution belongs to a Banach space $X^d(T)$ with larger $d$, then its energy displays higher localization.

We also observe that if
\begin{equation}
\label{til-Q}
\tilde{Q}_d=\sup_{0\leq t<T}\left\{\sum_{n=-\infty}^{+\infty}n^{2d}\left|q_n(t)\right|^2\right\}<\infty\ ,
\end{equation}
for some integer $d$, then
\[
\dis\sup_{0\leq t<T}\left\{\sum_{n=-\infty}^{+\infty}\left(1+n^2\right)^d\left|q_n(t)\right|^2\right\}<\infty\ ,
\]
because \mbox{$\tilde{Q}_{d_1}\leq\tilde{Q}_{d}$} for all $d_1<d$. This will allow us to simplify analysis by restricting consideration to $\tilde{Q}_d$ whenever convergence of sums will be questioned.

One can associate the map $\tilde{P}$ to the integral form of Eq. (\ref{int}):
\begin{equation}
\label{map}
\tilde{P}q_n(t)=p_n(t)+\sum_{m=-\infty}^{+\infty}\int_0^tG(n-m,t-s)F\left(q_m(s)\right) d s\ ,
\end{equation}
and respectively $Pq(t)=\left(\ldots,\tilde{P}q_{n-1}(t),\tilde{P}q_n(t),\tilde{P}q_{n+1}(t),\ldots\right) .$ Then a solution of (\ref{Seq1}) is nothing but a fixed point of the map $P$: \mbox{$q(t)=Pq(t)$}. This constitutes a basis for the use of the
Banach fixed point theorem which states that: if one has a contraction mapping $P$ from a closed subset $F$ of a Banach space $X^d$ onto $F$, then there exists a unique $q\in F$ such that $Pq=q$.  Thus, we have to define a closed subset of the Banach space as well as a functional that maps this subset into itself, the latter being a contraction. To this end we consider $R>0$ and $p(t)$ as in (\ref{func_q}), and define the following subset of $X^d(T)$:
\[
Y_R(T)=\left\{q\in X^d(T),\ \left\|q-p\right\|_{X^d}\leq R,\ q_n(0)=p_n(0)=a_n\right\}\ .
\]
Clearly, $Y_R$ is a closed subset of $X^d(T)$. Moreover, we observe that if $q\in Y_R(T)$, then $\left\|q\right\|_{X^d}\leq\left\|p\right\|_{X^d}+ R<\infty$, and $ºn^{2d}|q_n(t)|^2\leq\left\|q\right\|_{X^d}^2<\infty$.

Now we can formulate the main results of the paper, namely the existence theorems.

\smallskip

\begin{theorem}
\label{loc}
{\it (Local existence)} Assume that $a_n$ satisfies the condition (\ref{hp}), then there exists $T>0$ and a unique function $q(t)$, as in (\ref{func_q}), defined in $[0,T[$ which belongs to the Banach space $X^d(T)$, such that $q_n(t)$ satisfies (\ref{int}).
\end{theorem}

\smallskip

\begin{theorem}
\label{glob}
{\it (Global existence)}
The local solution of (\ref{int}) found in Theorem \ref{loc} can be extended for any $T>0$. Such global solution is unique.
\end{theorem}

\section{Proof of the local existence theorem}
\label{sec:local}

In this section we prove Theorem~\ref{loc}.  To shorten notations we suppress the subindex $X^d$ in the norm definition (i.e. $\|\cdot\|_{X^d}$ will be designated as $\|\cdot\|$), taking into account that no other norm will be used. Also we denote  constants (independent on the site number $n$ and time $t$) by $C$, whose values will however vary from line to line.

Let us consider some auxiliary results starting with
\begin{lemma}
\label{lemmaG}
For $t\in [0,T[$
\begin{equation}
\label{relG}
\left|G(0,t)\right|^2=1, \quad\mbox{and}\quad \left|G(n,t)\right|^2\leq\frac{K_j^2}{n^{2j}} \quad \mbox{for  $n\neq 0$, $\forall j\in \Zset^+$}\ ,
\end{equation}
where $K_j$ are constants depending on $j$ and $T$, and the series $\dis\sum_{n=-\infty}^{+\infty}n^{2d}|G(n,t)|^2$ is uniformly convergent.
\end{lemma}
\begin{proof}
For $n=0$ the estimate for the Green function trivially follows from its definition (\ref{Green}). For $n\neq 0$
integrating by parts we obtain:
\begin{equation*}
G(n,t)=\frac{i^j}{\pi\,2^j}\,\frac{1}{n^j}\int_{-\frac{\pi}{2}}^{\frac{\pi}{2}}f_{\sigma}^{(j)}(\sigma,t) d \sigma\ ,\ \ \  n\neq0\ ,
\end{equation*}
where $f_{\sigma}^{(j)}(\sigma,t)$ is the $j^{th}$ derivative of the function $f(\sigma,t)$, defined in (\ref{Green}), with respect to $\sigma$. The derivatives can be represented as:
\[
f_{\sigma}^{(j)}(\sigma,t)=\sum_{h=1}^j\phi_h(\sigma)t^he^{-4it\sin^2\sigma+2i\sigma
n}\ .
\]
Here the functions $\phi_h(\sigma)$, $h=1,\ldots,j$ are functions of $\sin(\sigma)$ and $\cos(\sigma)$, only, and thus $\phi_h(\sigma)$ are periodic bounded functions of $\sigma$. So we can deduce for $t\in [0,T[$:
\begin{equation*}
\left|G(n,t)\right|^2\leq\frac{1}{\pi^2\,(2n)^{2j}} \left(\int_{-\frac{\pi}{2}}^{\frac{\pi}{2}}\sum_{h=1}^j\left|\phi_h(\sigma)\right|t^h d \sigma\right)^2\leq \left(\frac{1}{2^j}\sum_{h=1}^jC_hT^{h}\right)^2\,\frac{1}{n^{2j}}\ ,
\end{equation*}
where $\dis C_h=\frac{1}{\pi^2}\int_{-\pi/2}^{\pi/2}|\phi_h(\sigma)|d\sigma$.
Now it is immediate to obtain (\ref{relG}) and derive the uniform convergence of the series: indeed, it is sufficient to take \mbox{$\dis K_j=\frac{1}{2^j}\sum_{h=1}^jC_hT^{h}$}, and $j>d$.
\end{proof}

It should to be mentioned here that the above result is nothing but a consequence of the Riemann-Lebesgue lemma (see e.g. \cite{Erd}) stating that 
\[
\int_{-\frac{\pi}{2}}^{\frac{\pi}{2}}f(\sigma, t) d \sigma=o\left(n^{1-N}\right)\ ,
\]
when $n\rightarrow\infty$,
i.e. that the integral decays faster that any power of $n$. Thus for all $d\geq 0$ we always can find $N$ sufficiently large such that the series   $\dis\sum_{n=-\infty}^{+\infty}n^{2d}|G(n,t)|^2$ is convergent.

From Lemma~\ref{lemmaG} we deduce a corollary important for the next consideration:
\begin{equation}
\label{ineq100}
\sum_{n=-\infty}^{\infty}n^{2d}|G(n-m,t)|^2\leq C_d\,(1+m^2)^{d}\ ,
\end{equation}
where $C_d$ is a constant depending on the power $d$.

For the next consideration it is also important that the localization principle holds also for the linear problem (\ref{Seq-lin}), what is guaranteed by the following
\begin{lemma}
\label{lem:lin}
The function:
\begin{equation}
\label{om}
p(t)=\left(\ldots,p_{n-1}(t),p_n(t),p_{n+1}(t),\ldots\right)\ ,
\end{equation}
where $p_n(t)$ is the solution for the linear problem (\ref{Seq-lin}), belongs to the Banach space $X^d(T)$, for any $T>0$, provided that (\ref{hp}) holds.
\end{lemma}
\begin{proof}
Using (\ref{relG}) with $j=d+1$, Cauchy-Schwarz inequality, and inequality (\ref{ineq100}) we obtain:
\begin{eqnarray*}
&& \sum_{n=-\infty}^{+\infty}n^{2d}\left|p_n(t)\right|^2=\sum_{n=-\infty}^{+\infty}n^{2d}\left|\sum_{m=-\infty}^{+\infty}a_mG(n-m,t)\right|^2\\
&& \leq  \sum_{m_1,m=-\infty}^{+\infty}|a_{m_1}|\cdot |a_{m}|\sum_{n=-\infty}^{+\infty} n^d\left|G(n-m_1,t)\right|\cdot n^d\left|G(n-m,t)\right|\\
&& \leq
\left[\sum_{m=-\infty}^{+\infty}|a_{m}|\left(\sum_{n=-\infty}^{+\infty}n^{2d}\left|G(n-m,t)\right|^2\right)^{1/2}\right]^2 \leq  C\left(\sum_{m=-\infty}^{+\infty}|a_m|m^d\right)^2\ .
\end{eqnarray*}
To conclude the proof we take into account that boundness of $A_d$ (see (\ref{hp})) implies convergence of the last sum.
\end{proof}

Now we are in a position to present the proof of the main Theorem~\ref{loc}:
\begin{proof}[Proof of Theorem \ref{loc}]
Let $q(t)\in Y_R(T)$, and $P$ is the map defined above. In order to prove that $Pq(t)\in Y_R(T)$ we have to show that $\left\|Pq-p\right\|\leq R$. Let
\[
w_n(t)\equiv Pq_n(t)-p_n(t)=\sum_{m=-\infty}^{+\infty}\int_0^tG(n-m,t-s)F\left(q_m(s)\right) d s\ .
\]
Thus we have to estimate (to shorten notations hereafter we use $\tau=t-s$ and $\tau_1=t-s_1$ and overbar hereafter stands for complex conjugation):
\begin{eqnarray*}
\sum_{n=-\infty}^{+\infty}n^{2d}|w_n|^2 &=& \sum_{n=-\infty}^{+\infty}n^{2d}\left|\sum_{m=-\infty}^{+\infty}\int_0^tG(n-m,\tau)\left|q_{m}(s)\right|^{2(p-1)}q_{m}(s) d s\right.\\
&\times&
\left.\sum_{m_1=-\infty}^{+\infty}\int_0^t\overline{G}(n-m_1,\tau_1) \left|q_{m_1}(s_1)\right|^{2(p-1)}\overline{q}_{m_1}(s_1) d s_1\right|\ .
\end{eqnarray*}

We will verify later on that the series which appear above are uniformly convergent so that we can exchange order of summation and integration and estimate:
\begin{eqnarray*}
&& \sum_{n=-\infty}^{+\infty}n^{2d}|w_n|^2 \leq\int_0^t d s  \int_0^t d s_1 \sum_{n=-\infty}^{+\infty}n^{2d}\\
&& \times \sum_{m=-\infty}^{+\infty}\n\left|q_{m}(s)\right|^{2p-1}\left|G(n-m,\tau)\right|\sum_{m_1=-\infty}^{+\infty}\n\left|q_{m_1}(s_1)\right|^{2p-1}\left|G(n-m_1,\tau_1)\right|\\
&&
\leq C\negthickspace\int_0^t\negthickspace d s  \int_0^t\negthickspace d s_1\sum_{m=-\infty}^{+\infty}(1+m^2)^d\left|q_{m}(s)\right|^{2p-1}\negthickspace\sum_{m_1=-\infty}^{+\infty}(1+m_1^2)^d \left|q_{m_1}(s_1)\right|^{2p-1}\\
&&
\leq C t^2 \|q\|^4N^{2p-3}\ .
\end{eqnarray*}
To obtain the last inequality we used (\ref{ineq100}). So we can conclude that
\[
\|w\|^2  \leq C\|q\|^{4} N^{2p-3}T^2
\leq  C\left(\|p\| +R\right)^4 N^{2p-3}T^2\ .
\]

Therefore we have only to choose $T>0$ sufficiently small, in such a way that:
\begin{equation}
\label{rel.1}
T<\frac{R}{C\left(\|p\| +R\right)^{4} N^{p-3/2}}\ .
\end{equation}
Then $\|Pq-p\|\leq R$, that is the functional $P$ maps $Y_R(T)$ into itself.

Now we will prove that $P$ is a contraction, that is, there exists a real number $\alpha$, $0<\alpha<1$, such that:
\[
\left\|Pq-P\tilde{q}\right\|\leq\alpha\left\|q-\tilde{q}\right\|\, \ \ \forall q,\tilde{q}\in Y_R(T)\ .
\]
To this end we define a function
\[
\Theta_m(t)\equiv\left|q_{m}(t)\right|^{2(p-1)}q_{m}(t)-\left|\q_{m}(t)\right|^{2(p-1)}\q_{m}(t)\ ,
\]
and estimate
\begin{eqnarray*}
&&\sum_{n=-\infty}^{+\infty}n^{2d}\left|Pq_n-P\q_n\right|^2=\sum_{n=-\infty}^{+\infty}n^{2d}\left|\sum_{m=-\infty}^{+\infty}\int_0^tG(n-m,\tau)\Theta_{m}(s) d s\right.\\
&&\times
\left.\sum_{m_1=-\infty}^{+\infty}\int_0^t\overline{G}(n-m_1,\tau_1) \overline{\Theta}_{m_1}(s_1) d s_1\right|\\
&&\leq
\int_0^t d s\int_0^t d s_1\sum_{n=-\infty}^{+\infty}n^{2d}\sum_{m=-\infty}^{+\infty}\left|\Theta_{m}(s)\right|\left|G(n-m,\tau)\right|\\
&&
\times\sum_{m_1=-\infty}^{+\infty}\left|\Theta_{m_1}(s_1)\right|\left|G(n-m_1,\tau_1)\right|\ .
\end{eqnarray*}
Likewise in the previous part we have exchanged order of summation and integration.

Next we notice that
\begin{equation}
\label{disq}
\left|\Theta_m(t)\right|\leq\left|q_m(t)-\q_m(t)\right|\sum_{k=0}^{2p-2}|q_{m}(t)|^{2p-2-k}|\q_m(t)|^{k}\ ,
\end{equation}
and thus
\begin{eqnarray*}
&& \sum_{m=-\infty}^{+\infty}\left(1+m^2\right)^{d+2}\Theta_m^2
\leq \\ &&
 \leq \sum_{m=-\infty}^{+\infty}\left(1+m^2\right)^{d+2}\left|q_m-\q_m\right|^2
\left(|q_m|+|\q_m|\right)^{4p-4}\\
&&
\leq \|q-\q\|^2\left(\|q\|+\|\q\|\right)^{4}  \left(N^{1/2}+\tilde{N}^{1/2}\right)^{4p-8}\ ,
\end{eqnarray*}
where $\dis\tilde{N}=\sum_{m=-\infty}^{\infty}|\tilde{q}_m|^2$.
Thus we obtain the following estimate
\begin{eqnarray*}
&&\sum_{n=-\infty}^{+\infty}n^{2d}\sum_{m=-\infty}^{+\infty}\left|\Theta_{m}(s)\right|\left|G(n-m,\tau)\right| \sum_{m_1=-\infty}^{+\infty}\left|\Theta_{m_1}(s_1)\right|\left|G(n-m_1,\tau_1)\right|\\
&&\sum_{m=-\infty}^{+\infty}\sum_{m_1=-\infty}^{+\infty}\left|\Theta_{m}(s)\right|\,\left|\Theta_{m_1}(s_1)\right|
\left(
\sum_{n=-\infty}^{+\infty}n^{2d}\left|G(n-m,\tau)\right|^2
\right)^{1/2}\\
&&\times
\left(
\sum_{n=-\infty}^{+\infty}n^{2d}\left|G(n-m_1,\tau_1)\right|^2
\right)^{1/2}\\
&&\leq
C\sum_{m=-\infty}^{+\infty}\sum_{m_1=-\infty}^{+\infty}(1+m^{2})^{d/2}(1+m_1^{2})^{d/2}
\left|\Theta_{m}(s)\right|\,\left|\Theta_{m_1}(s_1)\right|\\
&&\leq C\|q-\q\|^2\left(\|q\|+\|\q\|\right)^{4}\ ,
\end{eqnarray*}
where (\ref{relG}) has been used. From the last result we achieve:
\begin{eqnarray*}
\|Pq-P\q\|^2 &\leq& \sup_{0\leq t<T}\left\{\int_0^t d s \int_0^t d s_1C\left(\|q\|+\|\q\|\right)^{4}\|q-\q\|^2\right\}\\
& \leq &
C\left(\|p\|+R\right)^4\|q-\q\|^2T^2\ .
\end{eqnarray*}

Therefore given a positive number $\alpha$, such that $0<\alpha<1$ we can choose $T$ small enough so that $C\left(\|p\|+R\right)^4T< \alpha$, that is :
\begin{equation}
\label{rel.2}
T<\frac{\alpha}{C\left(\|p\|+R\right)^4}\ ,
\end{equation}
which implies that $P$ is a contraction in $Y_R$. Therefore choosing $T=T_0$ that satisfies the conditions (\ref{rel.1}), (\ref{rel.2}) we can conclude that there exists a unique fixed point $q\in Y_R(T)$ of the map $P$, i. e. $Pq=q$ for $0\leq t<T_0$
where $q(t)$ is given by (\ref{func_q}) and $q_n(t)$ satisfies (\ref{int}).
\end{proof}

One has to note that from the majorations achieved during the proof one deduces that the  series of functions are majorated by convergent numerical series. Then, from Weierstrass convergence criteria, one asserts that the series exhibit absolute convergence as well as uniform convergence, allowing one to change order of integration and summation.

\section{Global existence of solutions}
\label{sec:global}

In order to prove global existence we have to extend the solution (\ref{int}) for any $T>0$.

\begin{proof}
To this end we can use Zorn's lemma~\cite{zorn}, i.e. to extend the solution  to the maximal interval of existence which we denote by $[0,T_{\mathrm{max}}[$, and to prove that $T_{\mathrm{max}}=+\infty$. We proceed as follows: we suppose that $T_{\mathrm{max}}<+\infty$ and $\dis \|q\|\rightarrow+\infty\ $ as $\ T\rightarrow T_{\mathrm{max}}\ $ and then we will get a contradiction by an {\it a-priori} estimate. This is a trivial task since from inequality (\ref{ineq-Q}) we deduce:
\begin{equation}
\|q\|^2=\sup_{0\leq t<T}Q_d(t)\leq\sup_{0\leq t<T}\left\{Q_d(0)e^{C_0t}\right\}=Q_d(0)e^{C_0T}<\infty\ .
\end{equation}
Thus $T_{\mathrm{max}}=+\infty$ and the solution (\ref{int}) is a global solution of problem (\ref{Seq1}), (\ref{Seq1_1}).

It remains to prove the uniqueness of the global solution. Let $r_n(t)$ and $s_n(t)$ be two solutions of the problem (\ref{Seq1}), (\ref{Seq1_1}) and define \mbox{$z_n(t)=r_n(t)-s_n(t)$}. We will prove that $z_n(t)=0\ , \forall t\geq0$, what means that problem (\ref{Seq1}), (\ref{Seq1_1}) have a unique solution. To this end we consider the following function:
\begin{equation}
\label{z}
Z(t)=\sum_{n=-\infty}^{+\infty}\left(1+n^2\right)^d|z_n(t)|^2\ ,
\end{equation}
and compute its derivative in the form $
\dot{Z}(t)= D_1+D_2$ where
\begin{eqnarray*}
D_1&=&-2\Im\left\{\sum_{n=-\infty}^{+\infty}\left(1+n^2\right)^d
\big[z_{n+1}(t)\bar{z}_n(t)+z_{n-1}(t)\bar{z}_n(t)\big]\right\}\ ,
\\
D_2&=&-2\Im\left\{\sum_{n=-\infty}^{+\infty}\left(1+n^2\right)^d\bar{r}_n(t)s_n(t)\left(|r_n(t)|^{2(p-1)}-|s_n(t)|^{2(p-1)}\right)\right\}\ .
\end{eqnarray*}
With a similar argument used to estimate $Q_d(t)$ we obtain $D_1\leq C_0Z(t)$. Next, using that $|\Im (\bar{r}_ns_n)|\leq |r_n|\,|r_n-s_n|$, $|r_n|^2<N$, and $|s_n|^2<N$, we estimate
\begin{eqnarray*}
D_2&\leq& 2\sum_{n=-\infty}^{+\infty}\left(1+n^2\right)^d|\Im (\bar{r}_ns_n)|
|r_n-s_n|\left(|r_n|^{2p-3}+\cdots +|s_n|^{2p-3}\right)\\
&\leq& 2(2p-2)N^{2p-2}Z(t)\ .
\end{eqnarray*}
Therefore $Z(t)$ satisfies the following relation:
\[
\dot{Z}(t)\leq C_0Z(t)-4(p-1)N^{2p-2}Z(t)=CZ(t)\ ,
\]
and due to Gronwall's Lemma we have:
\[
0\leq Z(t)\leq Z(0)e^{Ct}\ .
\]
Since $Z(0)=0$ ($r_n(0)=s_n(0)=a_n$) we thus have that $Z(t)=0\ , \forall t\geq0$ and accordingly $z_n(t)= 0\ , \forall t\geq0$.
\end{proof}

\section{Generalizations}
\label{dissip}

\subsection{Dissipation}

The results obtained above are directly generalized to the DNLS equation with dissipation
\begin{equation}
\label{Seq1u}
i\dot{u}_{n}(t)+\Delta u_n(t)-\chi |u_n(t)|^{2(p-1)}u_n(t)+i\gamma u_n(t)=0\ , \qquad u_n(0)=a_n \ .
\end{equation}
where $\gamma$ is a positive constant characterizing dissipation. To this end we notice that by the means of the ansatz $u_n(t)=e^{-\gamma t}q_n(t)$ the problem (\ref{Seq1u}) can be reduced to:
\begin{equation}
\label{Seq2}
i\dot{q}_{n}(t)+\Delta q_n(t)-\chi e^{-2(p-1)\gamma t}|q_n(t)|^{2(p-1)}q_n(t)=0 \ ,
\qquad q_n(0)=a_n \ .
\end{equation}
The only change of the proofs presented above are in the definitions of $F$ (see (\ref{F})) which now reads
\begin{equation}
\label{FF}
F\left(q_n(t)\right)=e^{-2(p-1)\gamma t}\chi |q_n(t)|^{2(p-1)}q_n(t)\ .
\end{equation}
It is also important that in the case $\gamma>0$ one has $N(t)=e^{-\gamma t}N$, where $N$ is like in (\ref{N0}), and thus an estimate $ |q_n|^2\leq N$. Thus, without presenting a proof, we formulate the following theorem
\begin{theorem}
\label{diss}
Assume that $a_n$, $n\in\mathbb{Z}$, satisfy the condition (\ref{hp}), then there exists $T>0$ and a unique function $u(t)=\left(\ldots,u_{n-1}(t),u_n(t),u_{n+1}(t),\ldots\right)$, defined in $[0,T[$ which belongs to the Banach space $X^d(T)$, such that $u_n(t)$ satisy (\ref{int}). This solution (\ref{int}) can be extended for any $T>0$. Such global solution is unique.
\end{theorem}

\subsection{Long range interactions}

The results expressed in Theorems~\ref{loc} and \ref{glob} are essentially based on the properties of the Green function expressed by the formulas (\ref{relG}) and (\ref{ineq100}). Let us consider now a generalized discrete nonlinear Schr\"{o}dinger equation
\begin{equation}
\label{Seq100}
i\dot{q}_{n}(t)+\sum_{m=-\infty}^{\infty}K(n-m)q_m(t)-\chi |q_n(t)|^{2(p-1)}q_n(t)=0\ ,
\end{equation}
where long range interactions are included through the coupling constants $K(n-m)$ which are considered to be real, symmetric (i.e. $K(n)=K(-n)$), and decaying rapidly enough (i.e. such that $\dis\sum_{m=-\infty}^{\infty}|K(m)|<\infty$). As it is clear, in the case of Eq. (\ref{Seq1}) it was $K(n-m)=2\delta_{m,n}-\delta_{m,n+1}-\delta_{m,n-1}$.

The formulated problem (\ref{Seq100}) requires evident modification of the associated linear problem and results in a new Green function, which still can be written in a form (\ref{Green}) but now with the function
\begin{equation}
\label{Green1}
f(\sigma,t)\equiv e^{2i\sigma n-it\omega\left(\sigma\right)},\qquad \omega(\sigma)=1+2\sum_{m=1}^{\infty}K(m)\cos(2m\sigma)\ .
\end{equation}
Here $\omega(\sigma)$ is the dispersion relation of the lattice with long range interactions. The new defined function $f(\sigma,t)$ possesses the same properties as the $f(\sigma,t)$ defined in (\ref{Green}) which means that properties (\ref{relG}) and (\ref{ineq100}) hold also for (\ref{Green1}), and thus the analogs of the theorems ~\ref{loc} and \ref{glob} can immediately be formulated for the lattice with linear long-range interactions.

Coupling of the next-nearest neighbors can also be nonlinear, which modifies the nonlinearity of the Eq. (\ref{Seq1}) leading to a model
\begin{eqnarray}
\label{Seq102}
i\dot{q}_{n}(t)+\Delta q_n(t)-W_n(\dots,q_{n-1},q_n,q_{n+1},\dots,\bar{q}_{n-1},\bar{q}_n,\bar{q}_{n+1},\dots)=0\ ,
\end{eqnarray}
where $W_n(\cdot)$, $n\in\mathbb{Z}$, have the properties
\begin{eqnarray}
&&\nonumber W_n(\dots,q_{n-1}e^{i\varphi},q_ne^{i\varphi},q_{n+1}e^{i\varphi},\dots,\bar{q}_{n-1}e^{-i\varphi},\bar{q}_ne^{-i\varphi},\bar{q}_{n+1}e^{-i\varphi},\dots)\\
\label{w1}
&&\quad =e^{i\varphi}W_n(\dots,q_{n-1},q_n,q_{n+1},\dots,\bar{q}_{n-1},\bar{q}_n,\bar{q}_{n+1},\dots)\ ,
\\
&&\nonumber W_n(\dots,\lambda q_{n-m},\dots,\lambda q_{n+m},\dots,\lambda\bar{q}_{n-m},\dots,\lambda \bar{q}_{n+m},\dots)\\
\label{w2}
&&\quad =\lambda^{2(p-1)} W_n(\dots, q_{n-m},\dots, q_{n+m},\dots,\bar{q}_{n-m},\dots, \bar{q}_{n+m},\dots)\ ,
\end{eqnarray}
with $\lambda$ and $\varphi$ being an arbitrary real constants, and $p$ being a positive integer $p\geq 2$.

For the lattice (\ref{Seq102}), however, not all of the above results allow direct generalization. Indeed, one essential property used in the preceding sections was the conservation of the number of particles, i.e. $Q_0$ (or $N$, which is the same). But not all functions $W_n(\cdot)$ allow this conservation law. More precisely, in order to apply the above results of the localization of solution one has to impose one  more condition which reads
\begin{equation}
\label{cndw1}
\sum_{n=-\infty}^{\infty}\left[\bar{q}_nW_n(\dots,q_n,\dots,\bar{q}_n,\dots)-q_n\bar{W}_n(\dots,q_n,\dots,\bar{q}_n,\dots)\right]=0\ .
\end{equation}

In the same way, to formulate an analog of Theorems~\ref{loc} and \ref{glob} we have to impose one more constraint on $W_n(\cdot)$, namely
\begin{equation}
\label{cndw2}
\left|W_n(\ldots,q_n,\ldots,\bar{q}_n,\ldots)\right|\leq C|q_n|^{2(p-1)}\ .
\end{equation}

Then by analogy with the previous considerations we obtain the following
\begin{theorem}
\label{longr}
Assume that $a_n$, $n\in\mathbb{Z}$, satisfy the condition (\ref{hp}), moreover that (\ref{cndw1}) and (\ref{cndw2}) are valid. Then there exists $T>0$ and a unique function $q(t)=\left(\ldots,q_{n-1}(t),q_n(t),q_{n+1}(t),\ldots\right)$, defined in $[0,T[$ which belongs to the Banach space $X^d(T)$, such that $q_n(t)$ satisfy
\begin{equation}
q_n(t)=p_n(t)+\sum_{m=-\infty}^{+\infty}\int_0^tG(n-m,t-s)W_n(\dots,q_n(s),\dots, \bar{q}(s)_n,\dots)d s\ .
\end{equation}
This solution can be extended for any $T>0$. Such global solution is unique.
\end{theorem}

\subsection{Saturable nonlinearity}

As the last generalization of the above approach we consider a model with the saturable nonlinearity
\begin{equation}
\label{Seq103}
i\dot{q}_{n}(t)+\Delta q_n(t)-\chi\frac{|q_n(t)|^2q_n(t)}{A+B|q_n(t)|^2}=0\ ,
\end{equation}
where $A$ and $B$ are positive constants and $\chi$ is a real constant.
Evidently the nonlinear part of this last problem satisfies the conditions (\ref{cndw1}) and (\ref{cndw2}),  which allows us to formulate
\begin{theorem}
\label{sat}
Assume that $a_n$, $n\in\mathbb{Z}$, satisfy the condition (\ref{hp}), then there exists $T>0$ and a unique function $q(t)=\left(\ldots,q_{n-1}(t),q_n(t),q_{n+1}(t),\ldots\right)$, defined in $[0,T[$ which belongs to the Banach space $X^d(T)$, such that $q_n(t)$ satisfy
\begin{equation}
q_n(t)=p_n(t)+\chi\sum_{m=-\infty}^{+\infty}\int_0^tG(n-m,t-s)\frac{|q_n(s)|^2q_n(s)}{A+B|q_n(s)|^2} d s\ .
\end{equation}
This solution can be extended for any $T>0$. Such global solution is unique.
\end{theorem}

\section{Conclusion}

To conclude, we have proved that initially localized solutions of a number of models of the discrete nonlinear Schr\"{o}dinger type persist localized, with the same degree of localization, during an infinite time, and thus constitute a Banach space. The proof has been provided using the Banach fixed point theorem, which allowed us also to establish the well posedness of the Cauchy problem and thus the impossibility of existence of collapse. All these properties have rather transparent physical explanations which are respectively based on the spectrum of the underlining linear lattice having a cut-off frequency, and finite group velocity of the linear excitations (in the conventional continuous (nonlinear) Schr\"{o}dinger equation there is neither upper frequency limit nor bound on the speed of the linear excitations). Meanwhile application of the results was restricted to models which posses an integral of motion in the form of a number of particles $N$, which guaranteed boundness of $|u_n(t)|$ for all numbers of sites, $n$, and all times $t$. Although the most physically relevant models do have such an integral of motion, consideration of inter-site nonlinearities of more general types from the viewpoint of existence and localization of the solutions remains an open relevant problem.  It is also worth noting that other physically important generalizations of the obtained results are related to multidimensional and multi-atomic lattices.

\bigskip

\section{Acknowledgments}
P.P. thanks the Federal University of Rio de Janeiro for hospitality. The work of P.P has been supported by the ``Human Potential-Research Training Networks" fellowship (contract No. HPRN-CT-2000-00158). The cooperative work was supported by the Bilateral CAPES/GRICES program within the framework of Brazil/ Portugal agreement.


\begin{thebibliography}{99}

\bibitem{Vaz} {\em Localization and Energy Transfer in Nonlinear Systems}, Editors L. V\'azquez, R. S. MacKay, M. P. Zorzano (World Scientific, 2003); {\em Nonlinear Waves: Classical and Quantum Aspects}, Editors F. Kh. Abdullaev and V. V. Konotop (Kluwer Academic Publishers, 2004).

\bibitem{review}  D. Hennig, G. Tsironis, Phys. Reports {\bf 307}, 333 (1999);
P. G. Kevrekidis, K. \O . Rasmussen, A. R. Bishop, Int. J. Mod. Phys. {\bf B15}, 2833 (2001).



\bibitem{ILM} A. J. Sievers and S. Takeno, Phys. Rev. Lett. {\bf 61}, 970 (1988).

\bibitem{McKay1994}  R. S. MacKay, S. Aubry, Nonlinearity {\bf 7}, 1623 (1994);
D. Bambusi, Nonlinearity~{\bf 9}, 433 (1996).

\bibitem{ABK} G. L. Alfimov, V. A. Brazhnyi, V. V. Konotop, Physica  {\bf D 194}, 127 (2004).

\bibitem{Sulem} C. Sulem and P. L. Sulem, {\em The Nonlinear Schr\"{o}dinger Equation} (New York: Springer, 1999).

\bibitem{Bang} O. Bang, J. J. Rasmussen, and P. L. Christiansen, Physica {\bf D 68}, 169 (1993);
Nonlinearity {\bf 7}, 205 (1994).

\bibitem{MK} G. Perla Menzala, and V. V. Konotop, Applicable Analysis {\bf 75}, 157 (2000);
Nonlinear Analysis {\bf 54}, 261 (2003).

\bibitem{kant} L. V. Kantorovitch and G. Akilov, {\em Functional Analysis}, (Nauka, Moscow, 1977).

\bibitem{Erd} A. Erd\'elyi, {\em Asymptotic Expansions}, (Dover Publications, Inc., New York, 1956).

\bibitem{zorn} K. Junich, {\em Topology} (Springer-Verlag, New York, 1984).

\end{thebibliography}
\end{document}